\documentstyle[twocolumn]{mn}
\newcommand{\mincir}{\raise 
-2.truept\hbox{\rlap{\hbox{$\sim$}}\raise5.truept 
\hbox{$<$}\ }} 
\newcommand{\magcir}{\raise 
-2.truept\hbox{\rlap{\hbox{$\sim$}}\raise5.truept 
\hbox{$>$}\ }} 
\newcommand{\minmag}{\raise-2.truept\hbox{\rlap{\hbox{$<$}}\raise 
6.truept\hbox 
{$>$}\ }} 
\newcommand{\be}{\begin{equation}} 
\newcommand{\ee}{\end{equation}} 
\newcommand{\ba}{\begin{eqnarray}} 
\newcommand{\ea}{\end{eqnarray}} 
\newcommand{\brr}{\begin{array}} 
  
\newcommand{\err}{\end{array}} 
\newcommand{\bc}{\begin{center}} 
\newcommand{\ec}{\end{center}}

\newcommand{\ltapprox}{\raisebox{-0.5ex}{$\,\stackrel{<}{\scriptstyle\sim}\,$}}

\input{psfig.sty}  
 
\title{Effects of cluster galaxies on arc statistics} 
\author[Meneghetti et al.] 
  {Massimo Meneghetti$^{1,2}$, Micol Bolzonella$^3$, Matthias 
   Bartelmann$^2$, \newauthor Lauro Moscardini$^1$ and Giuseppe 
   Tormen$^1$ \\ 
   $^1$Dipartimento di Astronomia, Universit\`a di Padova, vicolo 
   dell'Osservatorio 5, I--35122 Padova, Italy\\ 
   $^2$Max-Planck-Institut f\"ur Astrophysik, P.O. Box 1523, D-85740 
   Garching, Germany\\ 
   $^3$Istituto di Fisica Cosmica G.P.S. Occhialini, via Bassini 15, 
   I-20133 Milano, Italy} 
 
\begin{document} 
 
\maketitle 
 
\begin{abstract} 

We present the results of a set of numerical simulations evaluating
the effect of cluster galaxies on arc statistics.

We perform a first set of gravitational lensing simulations using
three independent projections for each of nine different galaxy
clusters obtained from N-body simulations. The simulated clusters
consist of dark matter only. We add a population of galaxies to each
cluster, mimicking the observed luminosity function and the spatial
galaxy distribution, and repeat the lensing simulations including the
effects of cluster galaxies, which themselves act as individual
lenses. Each galaxy is represented by a spherical Navarro, Frenk \&
White (1997) density profile.

We consider the statistical distributions of the properties of the
gravitational arcs produced by our clusters with and without
galaxies. We find that the cluster galaxies do not introduce
perturbations strong enough to significantly change the number of arcs
and the distributions of lengths, widths, curvature radii and
length-to-width ratios of long arcs. We find some changes to the
distribution of short-arc properties in presence of cluster galaxies.
The differences appear in the distribution of curvature radii for arc
lengths smaller than $12''$, while the distributions of lengths,
widths and length-to-width ratios are significantly changed only for
arcs shorter than $4''$.

\end{abstract} 
 
 
\section{Introduction} 
 
Strong gravitational lensing by galaxy clusters distorts the images of
galaxies in the background of the clusters and thereby gives rise to
the formation of giant luminous arcs near the cluster
cores. Bartelmann et al.~(1998) showed recently that the statistics of
arcs is a potentially very sensitive probe for the cosmological matter
density parameter $\Omega_0$ and for the contribution $\Omega_\Lambda$
to the total density parameter due to the presence of the cosmological
constant. The reason can be summarised in form of three
statements. (i) Typical arc sources are located at redshifts around
unity. Lenses have to be placed approximately half-way between the
sources and the observer in order to be efficient, namely at redshifts
around $0.3$. (ii) The formation and evolution of galaxy clusters
depends strongly on the cosmological model (Richstone, Loeb \& Turner
1992; Bartelmann, Ehlers \& Schneider 1993). They tend to form only
recently in high-density universes and early in low-density
universes. In order to have a population of galaxy clusters capable of
forming arcs, at least a sufficiently large fraction of them must have
formed by redshift $\sim0.3$ because of argument (i). (iii) The
density of cluster cores depends on the formation redshift of the
clusters (Navarro, Frenk \& White 1996, 1997). The earlier a cluster
forms, the more compact it is. In universes with low density, clusters
therefore tend to be less compact the higher the cosmological constant
is.
 
Taken together, these arguments imply that more arcs can be observed
in a universe with low density and small cosmological constant: low
density makes clusters form earlier, and a low cosmological constant
makes them more compact individually.
 
This line of reasoning can easily be supported using Press-Schechter
(1974) theory. Numerical cluster simulations are necessary for
quantitative statements. They lead to the result that the number of
(suitably defined) large arcs on the whole sky is of order $10$ in an
Einstein-de Sitter universe, of order $100$ in a spatially-flat,
low-density universe with $\Omega_0=0.3$, and of order $1,000$ for an
open model with the same $\Omega_0$ and $\Omega_\Lambda=0$. The
observed number of arcs, which is of order $1,000$ when extrapolated
to the whole sky, then leads to the conclusion that $\Omega_0$ should
be low and $\Omega_\Lambda$ should be small or zero.
 
This result is important for several reasons. First, the effect is in 
principle easy to observe. It ``only'' requires to count arcs in a 
sufficiently large portion of the sky above a certain brightness, 
which are distorted by a certain minimum amount. Second, the effect is 
strong because order-of-magnitude differences are expected across 
certain popular cosmological models. Third, the consequence that 
$\Omega_\Lambda$ should be small or zero is at odds with the results 
obtained by the recent searches for supernovae of type Ia, which 
indicate that the universe is most likely spatially flat and has low 
density (Perlmutter et al.~1999).
 
The previous study (Bartelmann et al.~1998) neglected the granularity 
of the gravitational cluster potentials. It was ignored that the 
cluster galaxies could influence the lensing properties of 
clusters. Cluster galaxies have two principal effects. First, they 
tend to wiggle the critical curves of the cluster lenses, thereby 
increasing their lengths, and thus also the cross sections for strong 
lensing. Second, the larger local curvature of the critical curves can 
cause long arcs to split up, so that several shorter arcs can be 
formed where one long arc would have been formed in absence of the 
perturbing galaxies. These effects are counter-acting, and it requires 
numerical simulations to quantify their net effect. A secondary effect 
is that the local steepening of the density profile near cluster 
galaxies tends to make arcs thinner. 
 
We present in this paper numerical experiments to quantify the effect
of cluster galaxies on the cross sections for formation of large
arcs.
In particular, we address the question whether lensing by cluster
galaxies can invalidate the earlier result that clusters in a
high-density universe fail by two orders of magnitude to produce the
observed number of arcs.
We describe the cluster simulations and the technique used to
study the lensing properties of the clusters in Section~2. The
procedure for putting galaxies into them is presented in Section~3
where we also describe the method followed to compute the deflection
angles. The technique for identifying arcs and the results about the
distributions of their properties are detailed in Section~4. We finish
with a summary and a discussion in Section~5.
 
\section{The numerical method} 
 
\subsection{The cluster sample} 
  
The simulated clusters used as lenses in the present analysis are
those presented by Tormen, Bouchet \& White \shortcite{tormen97}. We
will only briefly describe them here and refer to that paper for more
details.
 
The sample is formed by the nine most massive clusters obtained in a
cosmological simulation of an Einstein-de Sitter universe, evolved
using a particle-particle-particle-mesh code. The initial conditions
have a scale-free power spectrum $P(k) \propto k^{-1}$, very close to
the behaviour of the standard Cold Dark Matter model on the scales
relevant for cluster formation. Despite the scale-free power spectrum,
a size can be assigned to the simulation box by determining the
variance of the dark-matter fluctuations and fixing the scale such
that the variance matches a certain value on that scale. Accordingly,
we demand that the {\em rms\/} density fluctuation in spheres of
radius $r=8\,h^{-1}\;$Mpc is $\sigma_8=0.63$, in rough agreement with
the normalisation of the power spectrum required to match the observed
local abundance of clusters \cite{white93}. The comoving size of the
simulation box then turns out to be $L=150\;$Mpc (in this paper a
Hubble constant of $50\,{\rm km\,s^{-1}\,Mpc^{-1}}$ is used).
 
Each cluster was obtained using a re-simulation technique (e.g.~Tormen
et al.~1997). In short, for each cluster in the cosmological
simulation new initial conditions were set up, in which the cluster
Lagrangian region was sampled by a higher number of particles than in
the original run. This allows a much higher spatial and mass
resolution in the re-simulated cluster. Particles not contributing to
the cluster were interpolated on a coarser distribution, which
describes the correct large-scale tidal field. These initial
conditions were evolved until the final time using a tree/SPH code
\cite{navarro93}, without gas, i.e.~as a pure N-body code.
 
Some of the cluster properties are summarised in
Table~\ref{clusprop}. Virial masses, encompassing an average
overdensity $\delta\rho/\rho=178$, range from $M_{\rm vir}=5.32 \times
10^{14}\,M_{\odot}$ to about $3 \times 10^{15}\,M_{\odot}$, while
one-dimensional {\em rms\/} velocities within the virial radius
$R_{\rm vir}$ range from 700 to 1300~km~s$^{-1}$. The average number
of particles within a cluster's $R_{\rm vir}$ is $\simeq 20,000$. The
gravitational softening $s$ imposed on small scales follows a cubic
spline profile, and was kept fixed in physical coordinates. Its value
is $s=20-25$~kpc at the final time, depending on the simulation.
Consequently, the force resolution in the simulations is $L/s \simeq
6,000$ to $7,500$ for the box and of the order of $R_{\rm
vir}/s\simeq100$ for each cluster.
 
\setcounter{table}{0} 
\begin{table} 
\centering 
\caption{Main properties of the simulated clusters. Column 1: cluster 
  name; column 2: average number of particles inside the virial 
  radius; column 3: virial mass; column 4: virial radius; column 5: 
  one-dimensional {\em rms\/} velocity within $R_{\rm vir}$.} 
\begin{tabular}{lrr@{$\times$}lrr} 
\hline\hline 
Cluster & 
$N_{\rm vir}$ & \multicolumn{2}{c}{$M_{\rm vir}$} & 
$R_{\rm vir}$ & $v_{\rm rms}$ \\ 
name & & \multicolumn{2}{c}{[$M_{\odot}$]} & [kpc] & 
[km s$^{-1}$] \\ 
\hline\hline 
 g15  & 39400 & $2.99$ & $10^{15}$ & 3870  & 1260\\ 
 g23  & 17400 & $6.76$ & $10^{14}$ & 2350  &  750\\ 
 g36  & 18200 & $1.51$ & $10^{15}$ & 3070  & 1000\\  
 g40  & 21300 & $5.32$ & $10^{14}$ & 2170  &  730\\  
 g51  & 23500 & $1.38$ & $10^{15}$ & 2990  & 1000\\  
 g57  & 24400 & $7.01$ & $10^{14}$ & 2380  &  780\\  
 g66  & 21400 & $1.10$ & $10^{15}$ & 2770  &  920\\  
 g81  & 14400 & $7.05$ & $10^{14}$ & 2390  &  750\\  
 g87  & 16200 & $6.21$ & $10^{14}$ & 2290  &  740\\  
\hline\hline 
\end{tabular} 
\label{clusprop} 
\end{table} 
  
\subsection{Lensing properties of the clusters} 
 
To study the lensing properties of a cluster, we first define a 
three-dimensional density field as follows. We centre the cluster in a 
cube of $6$~Mpc side length, which is covered with a regular grid of 
$N_{\rm g}=128^3$ cells. The resolution of the grid is therefore 
$R=6\,{\rm Mpc}/{N_{\rm g}^{1/3}}\approx47$~kpc. The three-dimensional 
cluster density $\rho$ at the grid points is calculated with the {\em 
Triangular Shape Cloud\/} method (see Hockney \& Eastwood 1988), 
which allows to avoid discontinuities. We extracted three different 
surface-density fields $\Sigma$ from each cluster by projecting $\rho$ 
along the three coordinate axes. This gives three lens planes per 
cluster which we consider independent cluster models for our present 
purpose. We can thus perform 27 different lensing simulations starting 
from our sample of nine clusters. 
 
We find the convergence $\kappa$ by dividing the projected density 
fields $\Sigma$ with the critical surface mass density for lensing,
defined as  
\begin{equation} 
  \Sigma_{\rm cr} \equiv 
    \frac{c^2} {4 \pi G}  
    \frac {D_{\rm S}}{D_{\rm L}D_{\rm LS}}\ , 
\label{nscr} 
\end{equation} 
where $c$ is the speed of light, $G$ is the gravitational constant and 
$D_{\rm L}$, $D_{\rm S}$ and $D_{\rm LS}$ are the angular-diameter 
distances between lens and observer, source and observer, and lens and 
source, respectively. We adopt $z_{\rm L}=0.4$ and $z_{\rm S}=2$ for 
the lens and source redshifts in the following. Hence, 
\begin{equation} 
  \Sigma_{\rm cr}\approx 2 \times 10^{15}\, 
  \frac{M_{\odot}}{{\rm Mpc}^2} \ . 
\end{equation}
Although real source galaxies are distributed in redshift, putting
them all at a single redshift is permissible because the critical
surface mass density (eq.~\ref{nscr}) changes only very little with source
redshift if the sources are substantially more distant than the
lensing cluster.
 
We further scale lens-plane coordinates with an arbitrary length
$\xi_0$, and source plane coordinates with the projected length
$\eta_0 \equiv D_{\rm S}/D_{\rm L}\,\xi_0$. The lens equation can then
be written
\begin{equation} 
  \vec{y}=\vec{x}-\alpha(\vec{x}) \ , 
\label{lenseq}  
\end{equation} 
with the reduced deflection angle 
\begin{equation} 
  \vec\alpha(\vec x)=\frac{1}{\pi}\,\int 
  \frac{\kappa(\vec x')\,(\vec x-\vec x')} 
  {|\vec x-\vec x'|^2}\,{\rm d}^2\vec x'\ . 
\label{angle} 
\end{equation} 
 
Aiming at the simulation of large arcs, we can concentrate on the 
cluster centres, i.e.~on the central regions of each lens plane. We 
therefore propagate a bundle of $1024\times1024$ light rays only 
through the central quarter of each lens plane. The corresponding 
resolution of this grid of rays is then fairly high, $\sim3$~kpc. 
 
The deflection angle $\vec{\alpha}_{ij}$ of each light ray ($i,j$) is
calculated with the discretised eq.~(\ref{angle}). Recall that the
convergence is defined on a grid, $\kappa_{mn}$, with
$m,n=1,\ldots,128$. We thus have
\begin{equation} 
  \vec{\alpha}_{ij}=\frac{1}{\pi}\, 
  \sum\limits_{m,n}\,\frac{\kappa_{mn}(\vec{x}_{ij}-\vec{x}_{mn})} 
  {|\vec{x}_{ij}-\vec{x}_{mn}|} \ , 
\end{equation} 
where $\vec{x}_{ij}$ and $\vec{x}_{mn}$ are the positions on the lens
plane of light ray ($i,j$) and grid point ($m,n$). The deflection
angles could diverge when the distance between a light ray and the
density grid-point is zero. Following Wambsganss, Cen \& Ostriker
\shortcite{wambsganss98}, we avoid this divergence by first
calculating the deflection angle on a regular grid of $128\times128$
``test rays'', shifted by half-cells in both directions with respect
to the grid on which the surface density is given. We then determine
the deflection angle of each light ray by bicubic interpolation
between the four nearest test rays.
 
The local properties of the lens mapping are described by the Jacobian
matrix of the lens equation (\ref{lenseq}),
\begin{equation} 
  A_{hk}(\vec{x}) \equiv \frac{\partial{y}_h}{\partial{x}_k} = 
  \delta_{hk}-\frac{\partial\alpha_h}{\partial x_k} \ . 
\label{jacobian} 
\end{equation} 
The shear components $\gamma_1$ and $\gamma_2$ are found from $A_{hk}$ 
through the standard relations 
\begin{eqnarray} 
  \gamma_1(\vec{x}) &=& 
  -\frac{1}{2}\left[A_{11}(\vec{x})-A_{22}(\vec{x})\right] \ , \\ 
  \gamma_2(\vec{x}) &=& 
  -\frac{1}{2}\left[A_{12}(\vec{x})+A_{21}(\vec{x})\right] \ , 
\label{kappagamsim} 
\end{eqnarray}   
and the magnification factor is given by the Jacobian determinant, 
\begin{equation} 
  \mu(\vec{x})=\frac{1}{\det A}= 
  [A_{11}(\vec{x})A_{22}(\vec{x})- 
   A_{12}(\vec{x})A_{21}(\vec{x})]^{-1} \ . 
\label{musim} 
\end{equation} 
 
Finally, the Jacobian determines the location of the critical curves
$\vec{x}_{\rm c}$ on the lens plane, which are defined by $\det A(\vec
x_{\rm c})=0$. Because of the finite grid resolution, we can only
approximately locate them by looking for pairs of adjacent cells with
opposite signs of $\det A$. The corresponding caustics $\vec{y}_{\rm
c}$ on the source plane are given through the lens equation,
\begin{equation} 
  \vec{y}_{\rm c}=\vec{x}_{\rm c}-\vec\alpha(\vec x_{\rm c}) \ . 
\label{cr} 
\end{equation} 
 
\section{Inserting cluster galaxies} 
 
The goal of this paper is to evaluate the effects of cluster galaxies
on arc statistics. For this purpose, we need to simulate a population
of galaxy lenses inside the cluster, in such a way that their
observational properties are well reproduced.
 
\subsection{Galaxy distribution} 
 
We start with the galaxy luminosity function. It is widely accepted
that the faint-end slope of the luminosity function depends on the
environment. In particular, it is steeper for cluster galaxies than
for field galaxies (Bernstein et al.~1995; De Propris et al.~1995).
Using a catalogue of isophotal magnitudes in the V band for 7,023
galaxies, Lobo et al.~\shortcite{lobo97} derived the luminosity
function in the Coma cluster in the magnitude range $13.5 < V \leq
21.0$ (corresponding to the absolute magnitude range $-22.24 < M_{V}
\leq -14.74$). Their results were fitted using both a steep Schechter
function,
\begin{equation} 
  S(M_V)=K_{\rm S}\,10^{0.4(\alpha+1)(M_V^*-M_V)} 
  \exp[-10^{0.4(M_V^*-M_V)}]  
\end{equation} 
(hereafter case ``S''), and a 4-parameter combination of a Schechter 
function with a Gaussian, 
\begin{equation} 
  G(M_V)=K_{\rm G}\,\exp[-(M_V-\mu)^2/(2\sigma^2)] \ , 
\end{equation} 
on its bright end (hereafter case ``S+G''). After excluding the three
brightest cluster members, they found $M^*_V=-29.0\pm2.0$ and
$\alpha=-1.59\pm0.02$ (with reduced $\chi^2=3.1$) in the ``S'' case,
while a much better fit (with $\chi^2=0.6$) was achieved in the
``S+G'' case, with parameters $M^*_V=-22.7\pm0.4$,
$\alpha=-1.80\pm0.05$, $\mu=-20.4\pm0.2$ and $\sigma=1.1\pm0.3$.

Notice that the Coma cluster has a mass similar to our simulated
clusters. In fact, Vedel \& Hartwick \shortcite{vedel98} recently
estimated that the mass of Coma within a radius of 5~Mpc falls in the
range $2.2-4.8 \times 10^{15} M_{\odot}$. We apply the two luminosity
functions from Lobo et al.~\shortcite{lobo97} to all our simulated
galaxy clusters.
 
Galaxy luminosities $L$ can be converted to baryonic masses $M$ with 
the relation 
\begin{equation} 
  \frac{M}{L}=4.0\,\left( 
    \frac{L}{4\times10^{10}\,L_{\odot}} 
  \right)^{0.35}\, 
  \left(\frac{M_{\odot}}{L_{\odot}}\right) \ , 
\end{equation} 
derived by van der Marel \shortcite{vandermarel91} from a study of a
sample of 37 elliptical galaxies. White et al.~\shortcite{white93}
found $\langle M/L \rangle = 3.2 (M_{\odot}/L_{\odot})$ by averaging
this relation over a luminosity function similar to that of the Coma
cluster.
 
Using the previous relations in Monte Carlo methods, we can now
generate a sample of galaxies with luminosities (and masses)
distributed like the galaxies in Coma. We will only consider galaxies
with a baryonic mass corresponding to the magnitude range where the
Lobo et al.~(1997) data were fit; this corresponds to a range $8.6
\times 10^8 \,M_\odot < M < 8.6 \times 10^{11}\,M_\odot$. In order to
be fully consistent with Lobo et al.~(1997), we include in our
simulated sample three more objects with luminosities (and masses)
corresponding to those of the three galaxies excluded from their
analysis.
 
The total number of galaxies to be placed in each simulated cluster is
determined by imposing a baryonic fraction as observed in Coma. We
adopt the estimate of White et al.~\shortcite{white93}, who found a
ratio between the baryonic mass in galaxies $M_{\rm b}$ and the total
mass of the cluster $M_{\rm tot}$ within the Abell radius $M_{\rm
b}/M_{\rm tot} \simeq 0.009$.
 
Finally, we must account for dark-matter haloes encompassing each
galaxy. We calculate total (virial) halo masses $M_{\rm vir}$ by
multiplying the baryonic masses $M_{\rm b}$ with the factor $f_{\rm
b}^{-1}$, where $f_{\rm b}$ is the average baryon fraction within
individual galaxies. Since this factor is not well known
observationally, we take a fiducial value of $f_{\rm b} \sim
5\%$. This number is close to the baryon fraction predicted by the
standard model of primordial nucleosynthesis.

\begin{figure*} 
{\centering \leavevmode  
\psfig{file=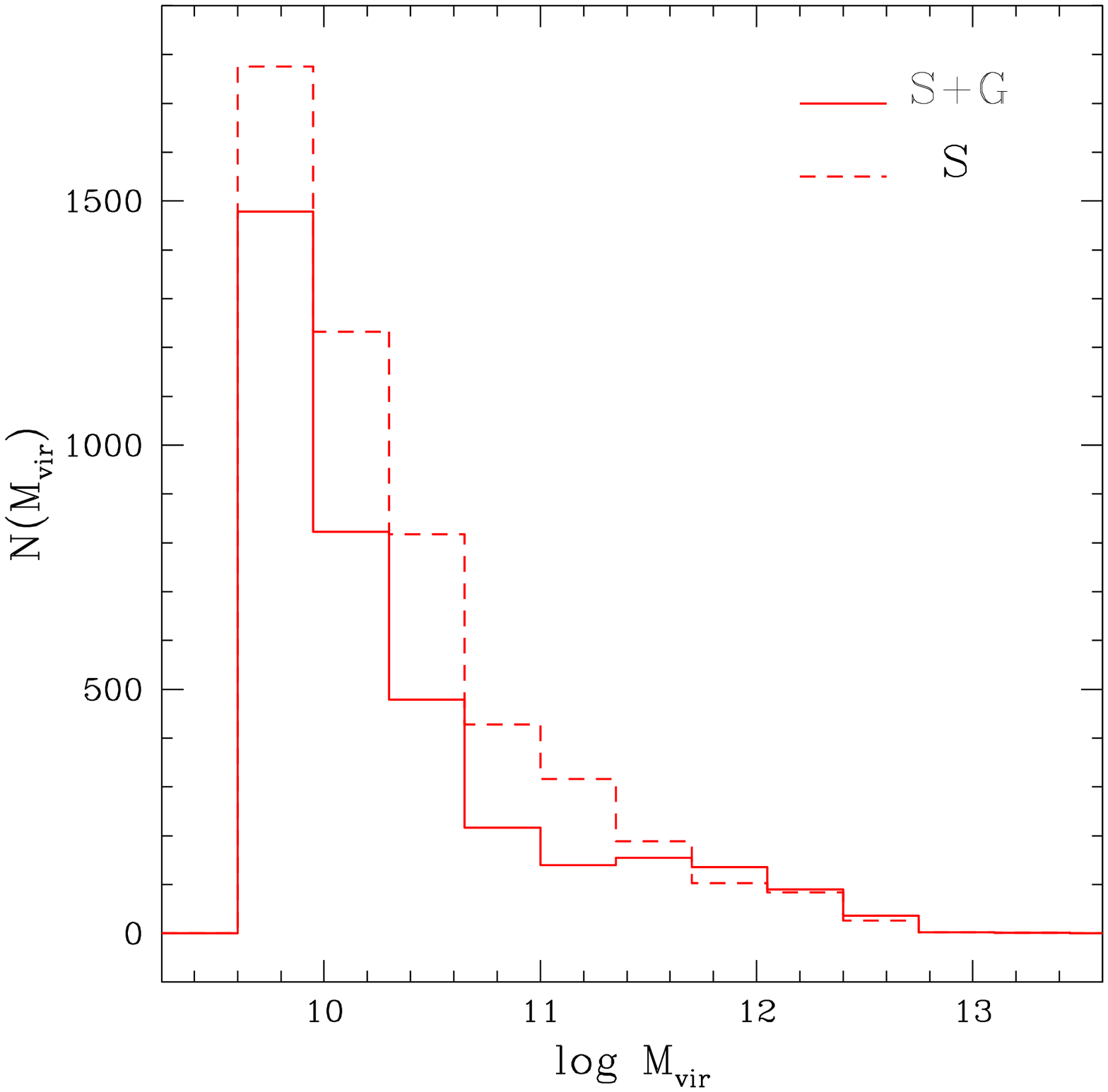,width=.49\textwidth} \hfil  
\psfig{file=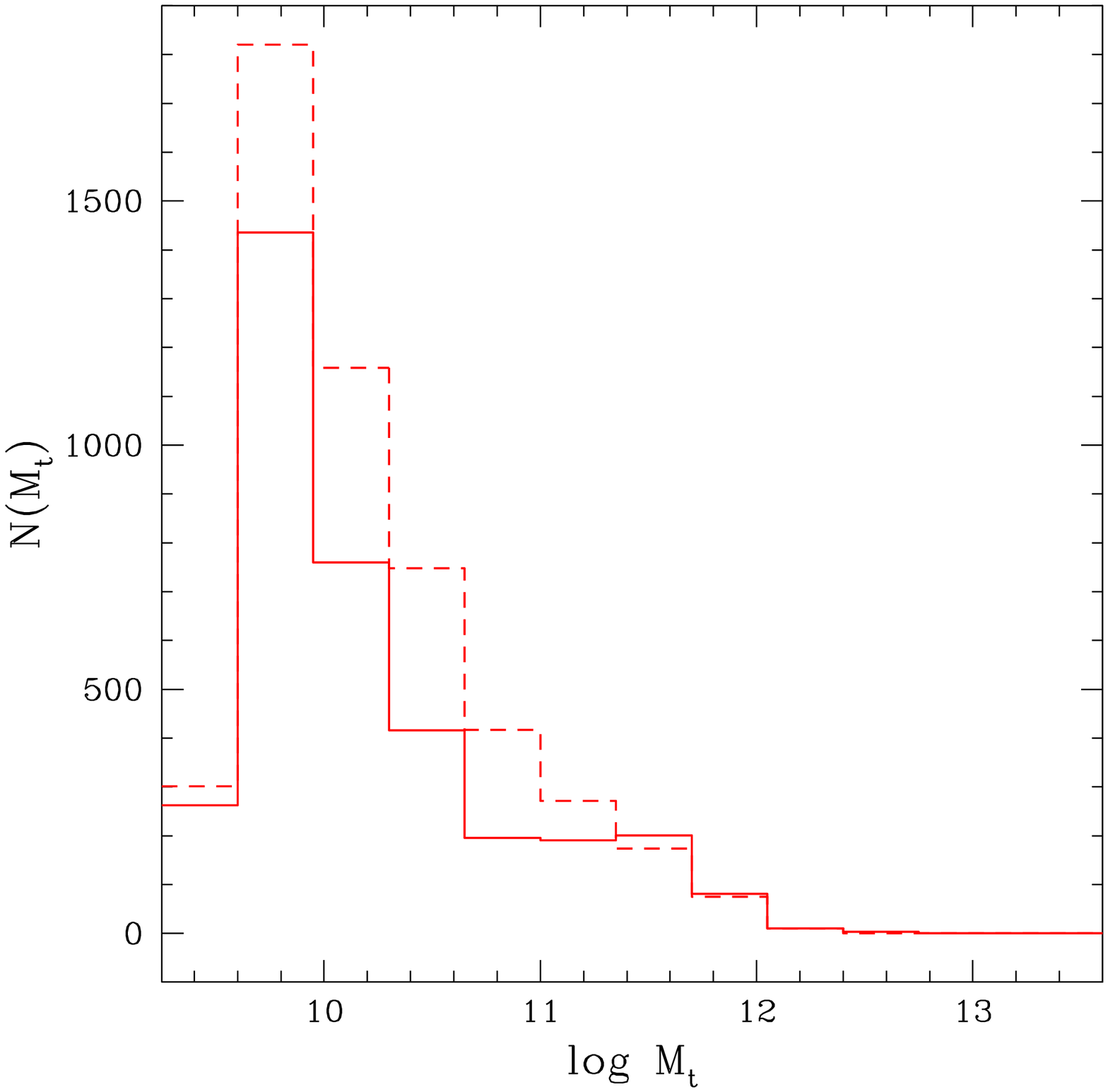,width=.49\textwidth} 
} 
\caption{Histograms of galaxy virial masses $M_{\rm vir}$ (left panel) 
  and of truncated masses $M_{\rm t}$ (right panel) for the simulated 
  sample of galaxies in cluster g15. Masses are in units of 
  $M_{\odot}$. The solid and dashed lines are for the ``S+G'' and 
  ``S'' luminosity functions, respectively.} 
\label{hist_masses} 
\end{figure*}

The right panel of Figure~\ref{hist_masses} shows the distribution of
galaxy virial masses $M_{\rm vir}$ in our most massive cluster (g15),
using both luminosity functions ``S+G'' and ``S'' with the best-fit
parameters from Lobo et al.~\shortcite{lobo97}. The obvious main
difference between the two cases is that the number of massive
galaxies is larger for the ``S+G'' luminosity function. The total
number of galaxies is therefore larger when the luminosity function
``S'' is adopted, where the same total mass is distributed over a
larger number of less massive galaxies.
 
Notice that the more massive a galaxy is, the stronger is the
perturbation caused by its potential. Therefore, the effect of cluster
galaxies on arc statistics will be more pronounced when the luminosity
function ``S+G'' is used.
  
We would like the galaxy number density to follow the mass density,
i.e.~galaxies should preferentially be positioned in overdense
regions. Furthermore, the most massive galaxies should be placed near
the centre of the cluster and in other large sublumps, i.e.~near the
highest density peaks. We achieve this by assigning to any given
position inside the cluster a probability for placing a galaxy which
is linearly proportional to the local density of dark matter.
 
Using a Monte Carlo technique, we first compute the number of galaxies
expected in each of the $32^3$ cells of a regular cubic grid covering
the cluster. Since the dark matter density was earlier defined on a
$128^3$ grid, each of these cells corresponds to $4^3$ cells of the
density grid, and contains a sufficient mass to host a galaxy.
 
We then sort the galaxy catalogue by decreasing mass, and start to
distribute them among the big cells with a probability proportional to
the local galaxy number density. This ensures that massive galaxies
are preferentially being placed in massive cells. Each galaxy is
finally randomly shifted within its grid cell.
 
\subsection{Galaxy mass profiles} 
 
Navarro, Frenk \& White \shortcite{nfw97} (hereafter NFW) found that
the equilibrium density profile of dark-matter haloes formed in
numerical simulations of several cosmological models (including most
CDM-like ones) is very well described by the radial function
\begin{equation} 
  \frac{\rho_{_{\rm NFW}}(r)}{\rho_{\rm cr}}= 
  \frac{\delta_{\rm c}}{(r/r_{\rm s})(1+r/r_{\rm s})^2} \ , 
\label{nfw} 
\end{equation} 
where $\delta_{\rm c}$ is a characteristic (dimension-less) density
and $r_{\rm s}$ is the scale radius; $\rho_{\rm cr}$ is the critical
density. We adopt this mildly singular density profile for our
simulated galaxies. Navarro et al.~\shortcite{nfw97} also showed that
the parameters $r_{\rm s}$ and $\delta_{\rm c}$ depend only on the
virial mass of a halo. Using the definition of the virial radius, they
are linked by the simple relation
\begin{equation} 
  \delta_{\rm c}=\frac{200}{3}\frac{k^3}{[\ln(1+k)-k/(1+k)]} \ , 
\label{cpar} 
\end{equation} 
where the halo concentration $k$ is defined as the ratio between the
virial radius $r_{\rm vir}$ and the scale radius $r_{\rm s}$.

We truncate each galaxy at a cut-off radius $r_{\rm t}$ where the 
galaxy density falls below the local cluster density. 
If we call $D$ the distance of the centre of the galactic halo from 
the cluster centre and $\rho$ is the cluster density profile,  
the cut-off radius $r_{\rm t}$ is determined by solving the equation 
\begin{equation} 
  \rho_{_{\rm NFW}}(r_{\rm t})=\rho(D-r_{\rm t}) \ . 
\end{equation} 
  
Knowing position and radius of each galaxy, we can determine what mass
must be subtracted from each small cell in exchange for the galaxy
mass. This is done via a Monte Carlo integration. The histogram of the
truncated galaxy masses $M_{\rm t}$ obtained for cluster g15
considering both luminosity functions is shown in the left panel of
Fig.~\ref{hist_masses}.
 
To avoid that all the galaxies concentrate near the cluster centre,
we impose that the cluster baryonic fraction is $\sim 0.9\%$ also
locally in each ``large'' cell. This implicitly means that we assume
that the relative concentration of mass and stars is similar in all
regions of the clusters without spatial segregation.
 
\begin{figure*} 
{\centering \leavevmode  
\psfig{file=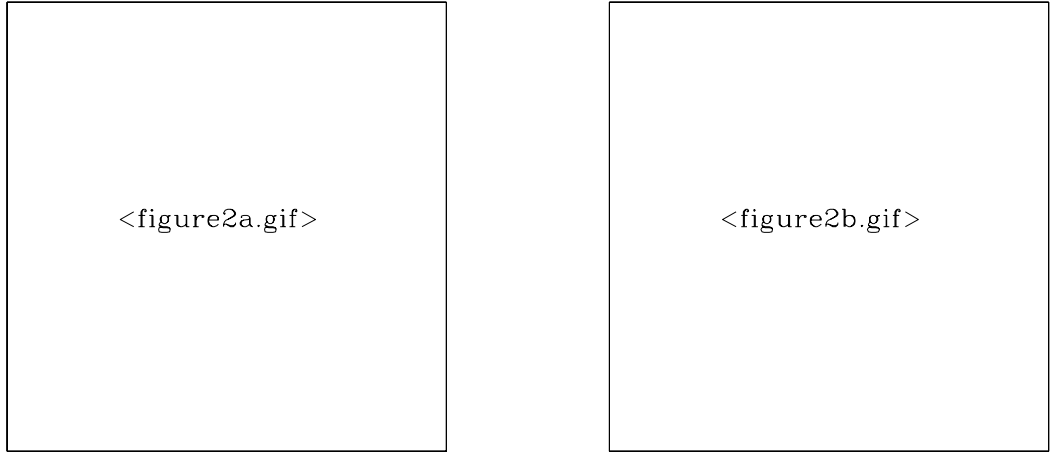,width=1.\textwidth}   
} 
\caption{Galaxy positions in cluster g15. Results for the two 
  different luminosity functions ``S+G'' and ``S'' are shown in the 
  left and right panels, respectively. Each galaxy is drawn as a 
  circle with radius equal to its truncated radius $r_{\rm t}$. Both 
  panels also show the convergence map. The scale of the figure is 
  3~Mpc on a side, which corresponds to an angular dimension of 
  $\simeq 333''$ for a cluster placed at $z_{\rm L}=0.4$. } 
\label{galmap} 
\end{figure*}

An example of a galaxy distribution obtained with this positioning
procedure is shown in Fig.~\ref{galmap}. All plots refer to cluster
g15 and show results for the two different luminosity functions
``S+G'' and ``S''. Each galaxy is represented in the figure by a
circle of radius $r_{\rm t}$, and it is superimposed on the cluster
convergence map, which is proportional to the surface mass
density. Galaxies near the cluster centre have a smaller radius
$r_{\rm t}$ because the cluster density is higher there, so only a
small part of the galaxy profile emerges above the cluster.
 
\begin{figure*} 
{\centering \leavevmode  
\psfig{file=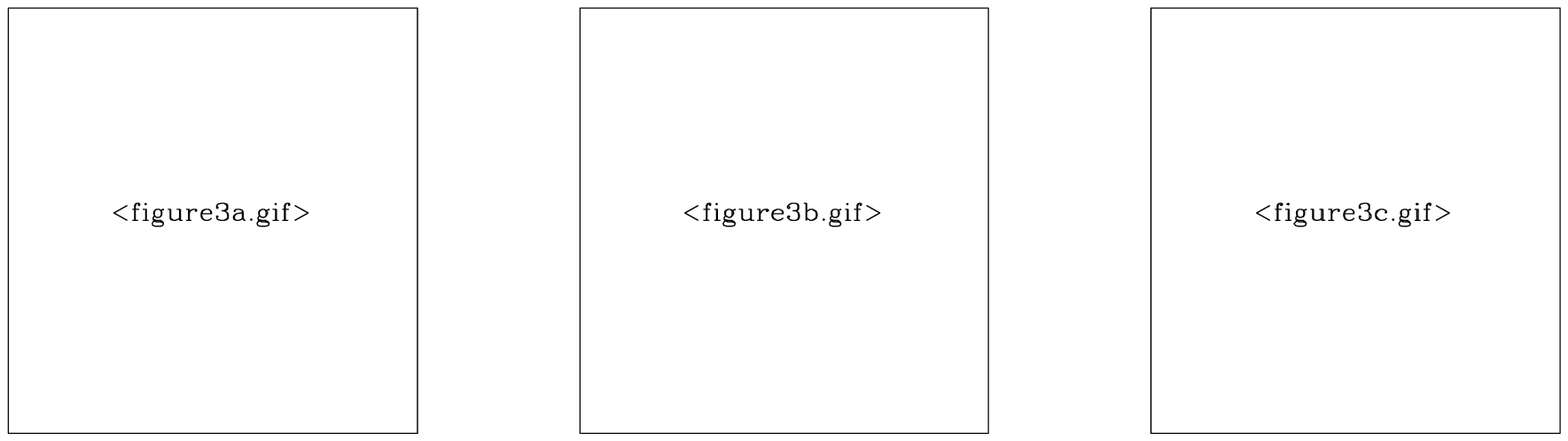,width=1.\textwidth}  
} 
\caption{Distribution of virial masses $M_{\rm vir}$ (left panel), 
  truncated masses $M_{\rm t}$ (central panel) and $M_{\rm
  t}/M_{\rm vir}$ ratios (right panel) for cluster g15, as a 
  function of distance $d$ from the cluster centre. The luminosity 
  function ``S+G'' was used. Each point represents a galaxy. Masses 
  are in units of solar masses. Median and (upper and lower) quartiles 
  of the distributions are shown by solid and dashed lines, 
  respectively.} 
\label{galsegr} 
\end{figure*}

Figure~\ref{galsegr} displays the distribution of virial and truncated
galaxy masses as a function of their distance from the cluster centre
$d$. In the right panel the distribution of the ratios $M_{\rm
t}/M_{\rm vir}$ is also presented. The median and the upper and lower
quartiles of the distributions are shown by solid and dashed lines,
respectively. The use of different luminosity functions does not
change the results significantly. For this reason, we will only show
results for the ``S+G'' luminosity function here and in the rest of
this paper, as it fits the Coma cluster data more accurately. The
plots show that the most massive galaxies are either near the cluster
centre or in secondary clumps at a distance of 2-3~Mpc. Notice that
the minimum truncated mass $M_{\rm t}$ is a slightly increasing
function of the distance $d$. In fact, even if the method is able to
place galaxies with the minimum allowed virial mass at all distances
from the centre, the density profile in the central part of the cluster
is so high that the emerging mass is strongly reduced. On the
contrary, at larger distance, $M_{\rm t}$ tends asymptotically to
$M_{\rm vir}$.

\subsection{Deflection angles} 
 
To compute the galaxy contribution to the deflection angles, we need
to calculate the projected NFW density profile of each galaxy.
Moreover, the profile needs to be truncated at the radius $r_{\rm t}$
previously defined. We can therefore distinguish between light rays
passing a galaxy inside or outside $r_{\rm t}$.
 
Defining $x$ in this subsection as the distance from the galaxy centre
in units of the scale radius $r_{\rm s}$ rather than $\xi_0$, the
convergence of the NFW profile (eq. \ref{nfw}) is
\begin{equation} 
  \kappa_{_{\rm NFW}}(x)=2\,\kappa_{\rm s}\,f(x) \ , 
\end{equation} 
with $f(x)$ given by eqs.~(\ref{app_f1},\ref{app_f2},\ref{app_f3}) in
the Appendix, and
\begin{equation} 
  \kappa_{\rm s}\equiv\rho_{\rm cr}\,\delta_{\rm c}\,r_{\rm s}\, 
  \Sigma_{\rm cr}^{-1} \ . 
\end{equation} 
The projected dimensionless mass within the radius $x$ is 
\begin{equation} 
  m(x)=4\kappa_{\rm s}\,\int_0^x x'f(x')dx' \ . 
\end{equation}  
 
The deflection angle due to a single galaxy on a light ray ($i,j$) 
passing the galaxy at a distance $x_{ij}$ is then 
\begin{equation} 
  \alpha_{ij}=\frac{m(x_{\rm min})}{x_{ij}} \ , 
\end{equation} 
where $x_{\rm min}\equiv\min(x_{ij},x_{\rm t})$ is the minimum of the 
scaled truncation radius $x_{\rm t}$ and the impact parameter 
$x_{ij}$. 
 
The total deflection angle of light ray ($i,j$) due to cluster 
galaxies is the sum of the contributions from all galaxies, 
\begin{equation} 
  \vec\alpha_{ij,{\rm gal}}=\sum\limits_{k=1}^{n_{\rm gal}}\, 
  \vec\alpha_{ij,n} \ , 
\end{equation}    
where $n_{\rm gal}$ is the total number of galaxies in the cluster. 
It must be added to the deflection angle due to the remaining dark 
matter in the cluster, $\alpha_{ij,{\rm dark}}$, to obtain the total 
deflection angle 
\begin{equation} 
  \vec\alpha_{ij}=\vec\alpha_{ij,{\rm dark}}+ 
  \vec\alpha_{ij,{\rm gal}} \ . 
\label{totangle} 
\end{equation} 
We are now able to compare the lensing effects of the clusters with
and without galaxies by inserting eq. (\ref{totangle}) into
eqs.~(\ref{lenseq}) and (\ref{jacobian}).
 
\section{Properties of the arc distribution} 
 
We now discuss the statistical properties of the arcs produced by the 
simulated clusters. We concentrate on some observable properties of 
the arcs, namely their lengths, widths, curvature radii and 
length-to-width ratios. 
 
The results obtained from the first set of 27 simulations using the
original simulated clusters (without galaxies) are compared to those
obtained after introducing galaxies as described in the previous
section. Hereafter, we will refer to the first type of simulations as
``DM'' (dark matter) simulations, and to the second type as ``GAL''
(galaxy) simulations.
 
\subsection{Identification of arcs and definition of their 
characteristics} 
 
As a first step, we need to find the images of a number of sources
sufficiently large for statistical analysis. We follow the method
introduced by Miralda-Escud\'e \shortcite{miralda93} and later adapted
to non-analytical models by Bartelmann \& Weiss
\shortcite{bartelmann94}. We refer to these papers for a more detailed
description of the method.
 
In the previous sections, the deflection angles were determined on a
grid of positions $\vec{x}_{ij}$ (with $1 \leq i,j \leq 1024$) in the
lens or image plane. Mapping these positions with the lens equation
(\ref{lenseq}), we obtain the source-plane coordinates
$\vec{y}_{ij}(\vec{x}_{ij})$ of the lens-plane grid points. Adopting
the terminology of Bartelmann \& Weiss \shortcite{bartelmann94}, we
call this discrete transformation a {\em mapping table\/}.
  
We model elliptical sources with axial ratios randomly drawn from the
interval $[0.5,1]$ and area equal to that of a circle of diameter
$d_{\rm s}=2''$. For numerical efficiency, we artificially increase
the probability of producing long arcs by placing a larger number of
sources near to or inside caustics, and a smaller number far away from
any caustics. Moreover, because of the convergence, only a restricted
part of the source plane can be reached by the light rays traced from
the observer through the lens plane. We then start with a coarse and
uniform grid of $32 \times 32$ sources defined in the central quarter
of the fraction of the source plane covered by the light rays
traced. Following Bartelmann \& Weiss \shortcite{bartelmann94}, we
double the source density and the resolution of the source grid where
the absolute magnification changes by more than unity across a grid
cell. The magnification at each point on the source plane can be found
from the mapping table. We repeat this procedure three times to obtain
the final list of sources. To give an example, we have $\simeq 5,000$
sources for the three projections of cluster g15.
 
From a statistical point of view, it is of course necessary to
compensate for the artificial increase in the number density of
sources near caustics. To do that, we assign a statistical weight of
$2^{2(N-n)}$ to each image of a sources placed during the $n$-th grid
refinement, where $N=3$ is the total number of refinements.
 
Given an extended source centred on $(y_1^{\rm c},y_2^{\rm c})$, we 
find all its images by searching the mapping table for points 
satisfying the condition 
\begin{equation} 
  \frac{(y_{1}-y_{1}^{\rm c})^2}{a^2}+ 
  \frac{(y_{2}-y_{2}^{\rm c})^2}{b^2} \leq 1 \ , 
\label{condiz} 
\end{equation} 
where $(y_{1},y_{2})$ are the components of the vector $\vec{y}$, and
$a$ and $b$ are the semi-axes of the ellipse representing the
source. We then use a standard {\em friends-of-friends\/} algorithm to
group {\em image points\/} within connected regions, since they belong
to the same image.

\begin{figure*}
{\centering \leavevmode 
\psfig{file=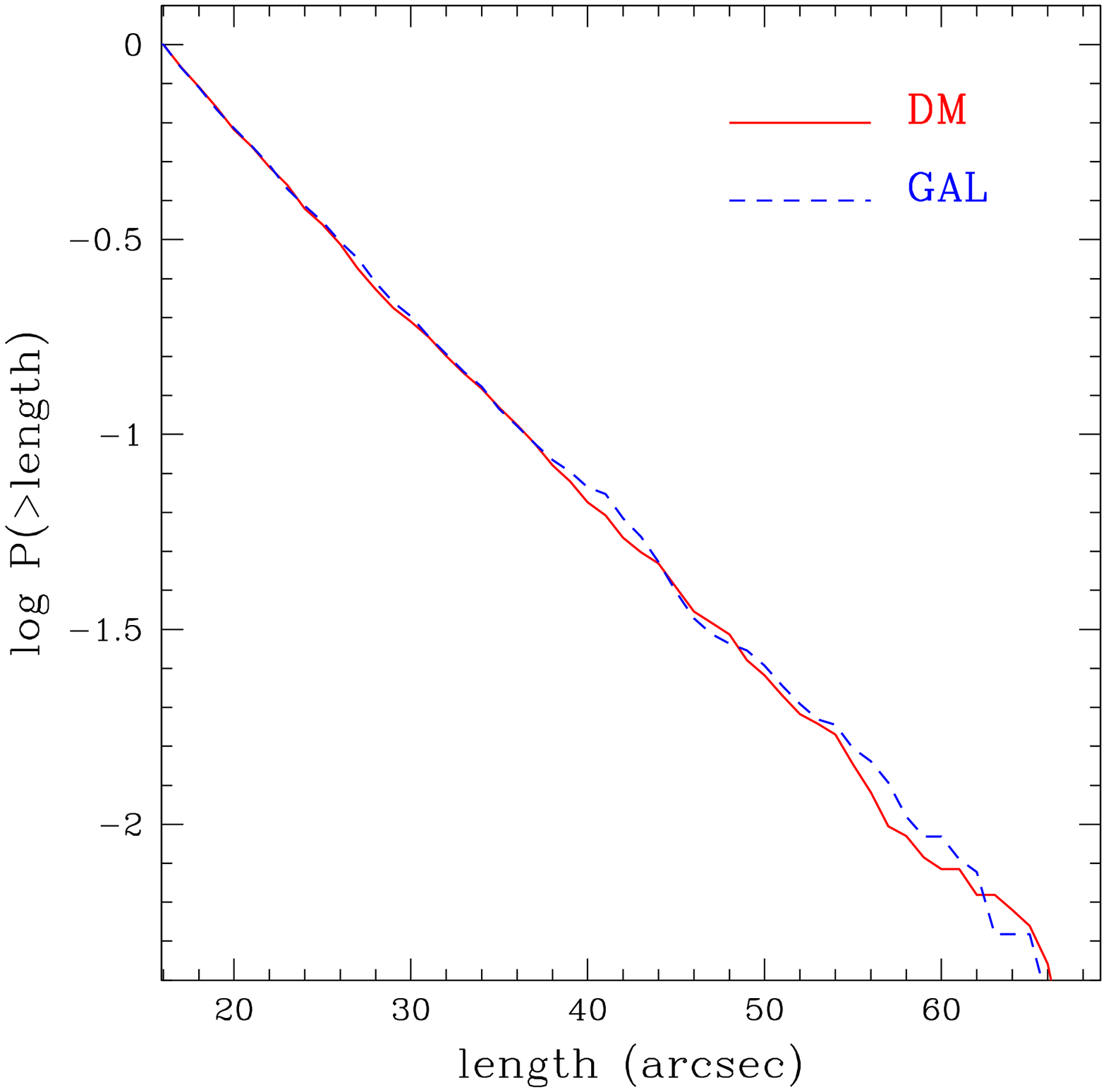,width=.49\textwidth} \hfil 
\psfig{file=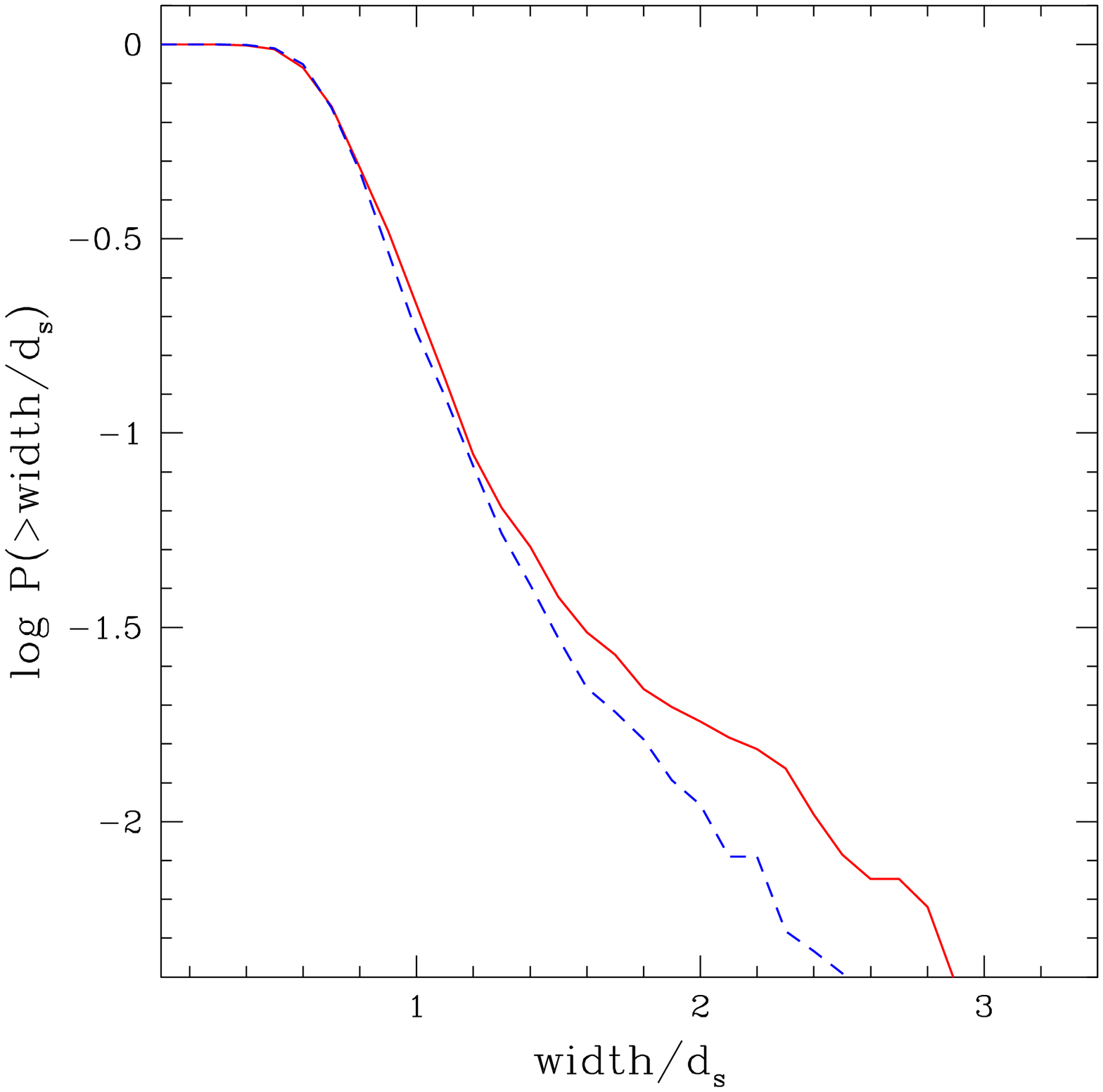,width=.49\textwidth}
}
\caption{The cumulative distributions for the lengths $l$ (in units of
arcsec) and the widths $w$ (in units of the source equivalent diameter
$d_{\rm s}=2''$) are presented in the left and right panels,
respectively. Results for the DM and GAL simulations are shown by
solid and dashed lines, respectively.
Typical bootstrap errors of the curves shown here grow from $\sim5\%$
for small arcs to $\sim15\%$ for large arcs.
}
\label{cumul}
\end{figure*}

The next step is the derivation of arc properties. We follow again the
method proposed by Miralda-Escud\'e \shortcite{miralda93} and
Bartelmann \& Weiss \shortcite{bartelmann94}. We define the area of
each image as the total number of image points. The circumference is
defined as the number of boundary points, i.e.~the number of image
points which are not completely enclosed by other image points.
 
We then find a circle crossing three image points, namely (a) its
centre, (b) the most distant boundary point from (a), and (c) the most
distant boundary point from (b). Since we use a grid on the image
plane, we cannot exactly find the centre of the image, and we have to
choose the image point which is mapped next to it. Notice that long
arcs can be merged from a few images, and there might exist more than
one image of the source centre. However, this is not a problem because
these points are located almost on the same circle.

We define the length $l$ and the curvature radius $r$ of the image through the
circle segment within points (b) and (c). To determine the image
width $w$, we search a simple geometrical figure with equal area and
length. For this fitting procedure, we consider ellipses, circles,
rectangles and rings. In the various cases, the image width is
approximated by the minor axis of the ellipse, the radius of the
circle, the smaller side of the rectangle, or the width of the ring,
respectively. A possible test for the quality of the geometrical fit
is given by the agreement between the circumferences of the
geometrical figure and the image.
 
\subsection{Results} 

\subsubsection{General properties}

For our statistical analysis we can use the results of 27 DM
simulations and the same number of GAL simulations produced by the
three projections along the Cartesian axes of nine clusters, which are
quite different, both in masses and shapes. As can be seen from Fig.~2
of Tormen et al. \shortcite{tormen97}, some of the structures are more
relaxed and show only one central density peak. In other cases the
clusters have evident substructures, and they are still in a dynamical
phase. Some of their lensing characteristics depend on these
properties. Regardless of the presence or absence of cluster galaxies,
the most massive clusters are the strongest lenses, as expected. For
example, the number of giant arcs (hereafter defined as the arcs
having a length larger than $16''$) produced by the g15 cluster
(having a mass of $\simeq 3 \times 10^{15} M_\odot$) is almost a
factor 8 larger than the number of those produced by the less massive
cluster of our sample (g40, with $M_{\rm vir}\simeq 5.3 \times 10^{14}
M_\odot$).

As already noticed by Bartelmann \& Weiss \shortcite{bartelmann94} and
Bartelmann, Steinmetz \& Weiss \shortcite{bsw95}, there is a strong
influence on arc statistics due to asymmetries and substructures in
the clusters. We find the same result in our sample (both in DM and
GAL simulations). The median of the distribution of the widths of
giant arcs is significatively smaller in the most compact projection
with respect to the other ones (e.g. $w\simeq 2.6''$ versus
$w\simeq 3.6''$ for cluster g81 and $w\simeq 3.2''$
versus $w\simeq 4.2''$ for cluster g40). On t he contrary, the
median of the length-to-width ratios is smaller for the projections
with a shallower density profile (e.g. $l/w\simeq 10 $ versus
$l/w\simeq 18 $ for cluster g81 and $l/w\simeq 9.5$ versus $l/w\simeq
15$ for cluster g40). This means that there is a larger probability to
have long and narrow arcs when the lens is more compact, i.e.~when the
central value of the convergence $\kappa$ is larger. Secondary
overdensities can also affect arc statistics. In fact, they produce a
shear field $\gamma$ which can change the shape of the critical lines,
defined as the curves on which $\det A=(1-\kappa)^2-\gamma^2=0$. We
find examples (e.g.~the cluster g15) in which large values of $\gamma$
can move the tangential critical lines to regions where the
convergence $\kappa$ is small.

\begin{figure}
{\centering \leavevmode 
\psfig{file=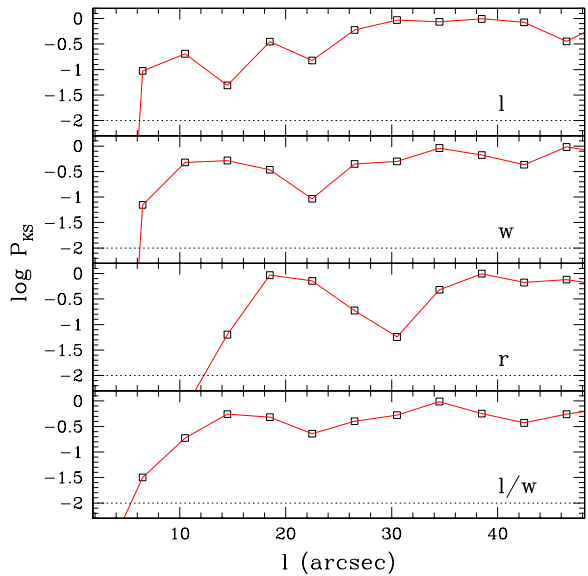,width=.49\textwidth} \hfil 
}
\caption{The behaviour of the probability $P_{\rm KS}$ (as computed in
a Kolmogorov-Smirnov test) that the arc property distributions in data
sets obtained from the simulations DM and GAL can be drawn from the
same parent distribution. Subsamples of arcs with a given length $l\pm
2''$ are considered. The panels refer to different properties: length
$l$, width $w$, curvature radius $r$ and length-to-width $l/w$ from
top to bottom. The dotted lines show the 1 per-cent level.}
\label{kstest}
\end{figure} 

\subsubsection{Distributions of the arc properties}

We now present the statistical analysis of the arc properties (length
$l$, width $w$, length-to-width ratio $l/w$ and curvature radius $r$).
We exclude from this statistical analysis all images represented by a
single grid-point, i.e.~produced by isolated rays: it would be
impossible to define the previous quantities for them.

As mentioned previously, we assign to each arc a weight depending on
the degree of the iteration in which the corresponding source was been
placed. In practice, the weight is proportional to the sky area which
is sampled by the source. The following distributions use this
normalisation.

The total number of arcs in the whole set of DM and GAL simulations is
quite similar: 447,112 and 451,782, respectively. The majority of
these arcs is quite short. Considering only giant arcs, defined as
arcs whose length is larger than $16''$, the sample reduces to 1,823
and 1,702 arcs for the DM and GAL simulations, respectively.

In Fig.~\ref{cumul}, we show the cumulative functions of lengths and
widths of these giant arcs. The length distributions (left panel) do
not seem to be sensitive to the cluster galaxies.  We found a similar
result also for the distributions of arc curvature radii (not shown in
the figure). On the other hand, the distributions of the arc widths
(right panel) show some differences between DM and GAL simulations, in
that the arcs are slightly thinner when galaxies are
included. Consequently, some small differences are also found in the
distributions of arc length-to-width ratios.

We checked by means of a bootstrapping analysis whether the
distribution functions of arc properties are affected by the relative
smallness of the cluster sample used. Across $10^5$ bootstrapped
samples constructed from the source distributions behind our 27
cluster fields, the {\em rms\/} scatter about the mean reached at most
$\sim15\%$. Bearing in mind that there are order-of-magnitude
differences between arc numbers expected in different cosmologies,
such uncertainties in the arc cross sections are entirely negligible.

Therefore, the previous results seem to indicate that the
characteristics of long arcs are only slightly changed by the presence
of galaxies. In order to evaluate whether the differences between the
two arc samples (DM vs.~GAL) depend on the arc length, we selected
subsamples of arcs with length in the range $l\pm\Delta l$ (with
$\Delta l=2''$). Then, by performing a Kolmogorov-Smirnov test, we
compare the arc property distributions in each subset of given arc
length. We show in Fig.~\ref{kstest} the significance level obtained
from the test as a function of $l$ for the four arc properties
considered here: length $l$, width $w$, curvature radius $r$ and
length-to-width ratio $l/w$. It can be seen from that plots that for
all arc properties the probability $P_{\rm KS}$ that data sets
obtained from the simulations DM and GAL can be drawn from the same
parent distributions becomes lower than $1\%$ only for very short
arcs. The differences are significant for $l\ltapprox 12''$ for the
curvature radii, and for $l\ltapprox 4''$ for the other properties. Once
again, these results indicate that giant arcs are generally not
significantly perturbed by cluster galaxies.

\begin{figure*}
{\centering \leavevmode 
\psfig{file=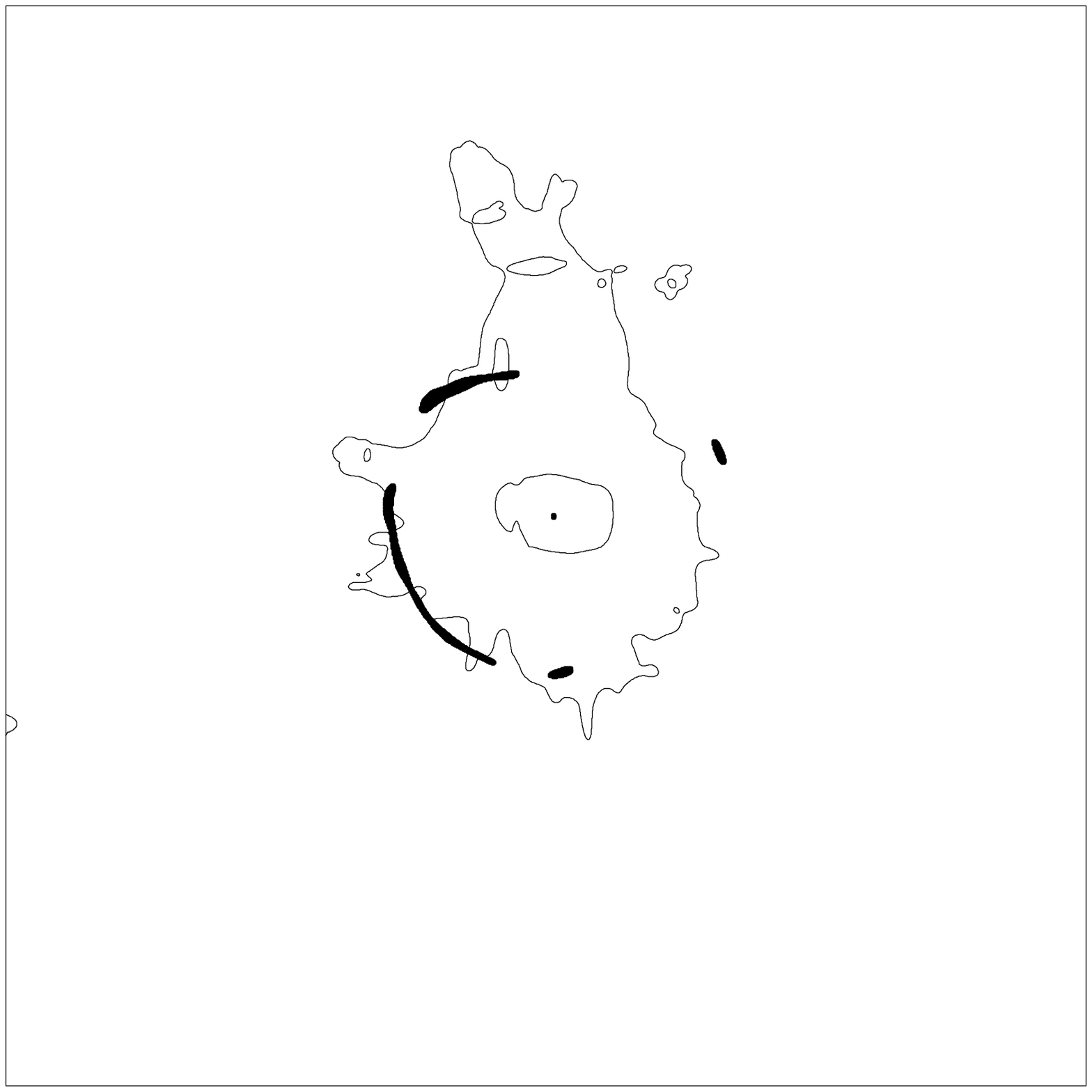,width=.49\textwidth} \hfil 
\psfig{file=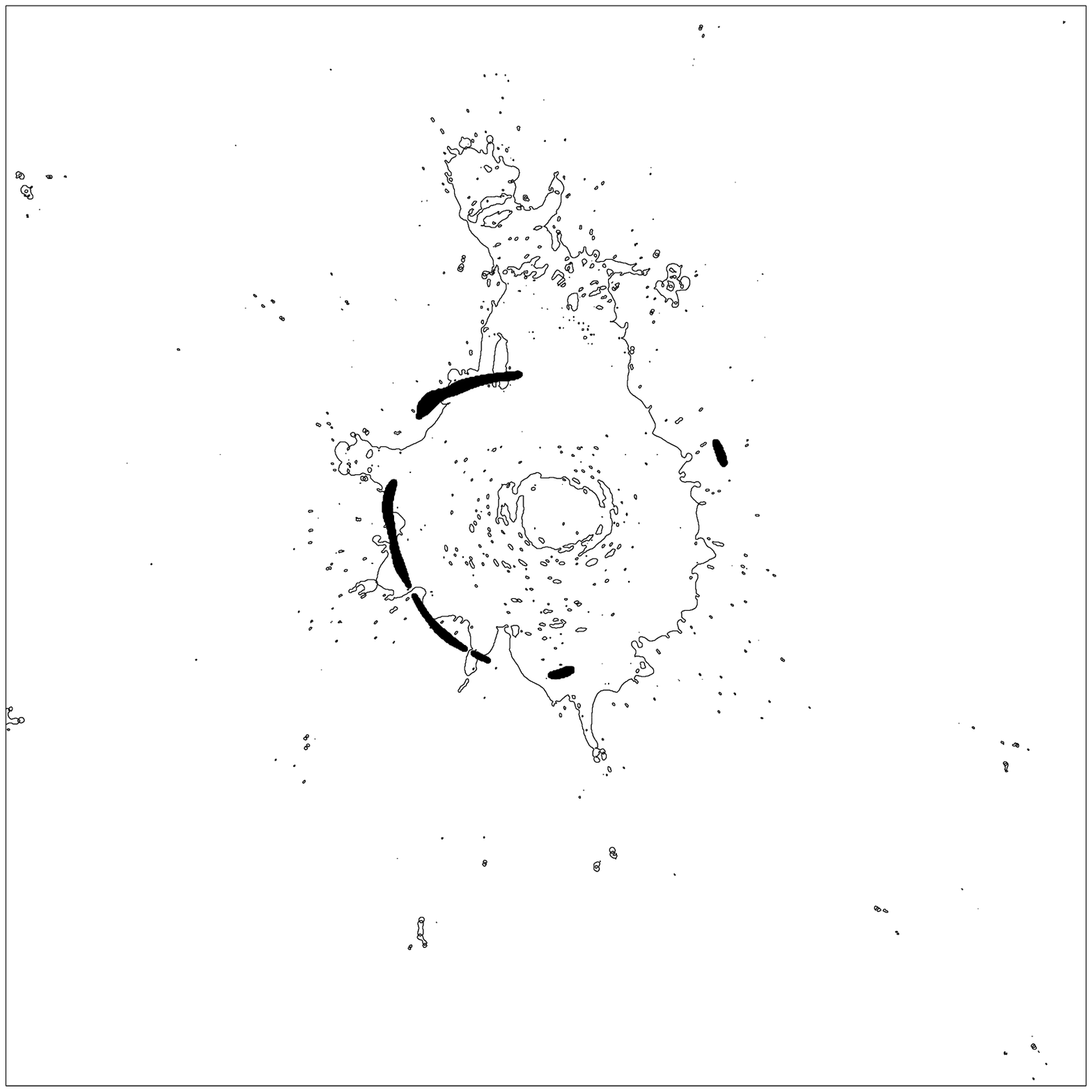,width=.49\textwidth}
}
\caption{Example of critical curves and arcs obtained in a DM (left
panel) and GAL simulations (right panel). The angular dimensions of
the plot correspond to $\simeq 333''$. It can be noticed that the
presence of galaxies acting as gravitational lenses can split long
arcs in several arclets, and increases the length of the critical
curves.}
\label{arcs}
\end{figure*}

\section{Discussion and Conclusions}

Three principal effects on the arc characteristics were expected due
to the presence of cluster galaxies. First, the cluster critical
curves wiggle around individual galaxies, increasing their length (see
the example shown in the right panel of Figure \ref{arcs}).  For this
reason, the cluster cross section for strong lensing would tend to be
increased and a larger number of long arcs would be expected.

At the same time, the curvature of the critical lines also increases.
Galaxies could therefore perturb some arcs and split them into several
shorter arclets, as can be seen for the longest arc in
Fig.~\ref{arcs}. Obviously, this effect acts such as to decrease the
cross section for strong lensing.

Finally, the local steepening of the density profile near cluster
galaxies tends to make arcs thinner.

The results of the Kolmogorov-Smirnov test indicate that the effect of
cluster galaxies is negligible if very short arcs are excluded from
the statistical analysis. This means than the first two effects
previously mentioned are almost exactly counter-acting, and the
splitting of some long arcs is compensated by the increased
strong-lensing ability of the clusters.

Moreover, considering all the giant arcs (larger than $16''$), the
galaxies tend to make them slightly thinner, as expected. However,
this effect is weak, indicating that the galaxies do not produce
perturbations strong enough to systematically affect all the arcs.

On the other hand, as confirmed by the Kolmogorov-Smirnov test, there
are some significant differences between the property distributions of
short arcs. These arcs do not form in the central regions of the
clusters, where most of the mass is concentrated and where long arcs
form instead. In such dense regions only a small fraction of the total
galaxy mass emerges from the underlying dark-matter distribution. This
does not happen in the outer regions of the clusters, where the dark
matter density is lower and the galaxies stick out almost completely
above the smooth cluster matter profile. For these reasons, the impact
of such galaxies is stronger, and several secondary short critical
lines form around them, as can be seen again in
Fig.~\ref{arcs}. Therefore, arcs forming far from the cluster centre
tend to be shorter and thinner, with larger curvature radii, and the
property distributions change significantly when the galaxies are
included in the simulations.
 
Bootstrap resampling of the 27 cluster fields shows that the {\em
rms\/} uncertainty of the cumulative arc distributions amounts to at
most $\sim15\%$, indicating that our cluster sample is large enough
for the results to be reliable.

These results allow us to conclude that the granularity of the cluster
potential due to the cluster galaxies has negligible effects on the
statistics of giant arcs
in an Einstein-de Sitter universe. What is more, we believe that we
can conclude that cluster galaxies also have negligible effects on
lensing by clusters in low-density universes. Such clusters form
earlier and are therefore more compact than those in an Einstein-de
Sitter universe. This implies that the strong-lensing cross sections
contributed by individual cluster galaxies are relatively even less
important compared to the cross sections of the clusters than in the
Einstein-de Sitter case. If galaxies have no effect on arc cross
sections under circumstances when the clusters themselves are the
weakest lenses, they will be entirely negligible when embedded into
stronger-lensing clusters.
This also means that previous predictions of the number of large arcs
produced by galaxy clusters via strong gravitational lensing, which
were obtained from numerically modelled clusters in which individual
galaxies are not resolved, can safely be used in comparisons with
observational data.

Our results are well compatible with a recent independent study by
Flores, Maller \& Primack (1999), who perturbed a pseudo-elliptical
cluster mass distribution with galaxies modelled as truncated
isothermal spheres and found only a negligible enhancement of the
smooth cluster's arc cross section.

\section*{Acknowledgements}
We are grateful to Francesco Lucchin for useful discussions. This work
was partially supported by Italian MURST, CNR and ASI, by the
Sonderforschungsbereich 375 for Astro-Particle Physics of the Deutsche
Forschungsgemeinschaft and by the TMR european network ``The Formation
and Evolution of Galaxies" under contract n. ERBFMRX-CT96-086. MM, LM,
GT thank the Max-Planck-Institut f\"ur Astrophysik for its hospitality
during the visits when this work was completed. MM acknowledges the
Italian CNAA for financial support.

\appendix
\section{Projected density profile of the galaxies}

We present in this appendix the detailed formulae for the projection
of the NFW density profile eq.~(\ref{nfw}) for the cluster galaxies.
  
Considering a galaxy with a truncation radius $r_{\rm t}$, the projected
NFW density profile is given by
\begin{equation}
  \Sigma _{_{\rm NFW}}(\xi)=2\int_0^{z_{\rm max}}
  \rho_{_{\rm NFW}}(r)\,{\rm d}z\ ,
\label{prnf}
\end{equation} 
where $z$ is the coordinate along the line of sight and $\xi$ is the
component of $r$ perpendicular to $z$. The maximum of $z$ is given by
$z_{\rm max}=\sqrt{r_{\rm t}^2 - \xi^2}$.

Using the dimensionless coordinate on the projection plane $x \equiv
\xi/r_s$ and defining the quantities $u \equiv \mbox{arcsinh} (z/\xi)$
and $\kappa_{\rm s}\equiv \delta_{\rm c} \rho_{\rm cr}\Sigma_{\rm
cr}^{-1}$, the previous equation can be written as
\begin{equation}
  \Sigma_{_{\rm NFW}}(x)=2\kappa_{\rm s} \Sigma_{\rm cr} f(x) \ ,
\end{equation}
where 
\begin{eqnarray}
  f(x) &=& -\frac{2}{(x^2-1)^{3/2}}
  \arctan\left[\frac{x-1}{\sqrt{x^2-1}}
  \tanh\left(\frac{u}{2}\right)\right] \nonumber \\
  &+& \left.\frac{1}{x^2-1}\frac{x\sinh u}{1+x\cosh u}
  \right|_0^{u_{\rm max}} 
\label{app_f1}
\end{eqnarray}
if $ x>1$;
\begin{equation}
  f(x)=\frac{2\cosh(\frac{u}{2})\sinh(\frac{u}{2})}{3(1+\cosh u)^2}+
  \left.
  \frac{4\cosh (\frac{u}{2})^3\sinh(\frac{u}{2})}{3(1+\cosh u)^2}
  \right|_0^{u_{\rm max}} 
\label{app_f2}
\end{equation}
if $x=1$; and
\begin{eqnarray}
  f(x) &=& \frac{2}{{1-x^2}^{3/2}}
  \mbox{arctanh}\left[\frac{1-x}{\sqrt{1-x^2}}
  \tanh\left(\frac{u}{2}\right)\right] \nonumber \\
  &+& \left.\frac{1}{1-x^2} \frac{x \sinh u}{1+x \cosh u}
  \right|_0^{u_{\rm max}}
\label{app_f3}
\end{eqnarray}
if $x<1$. 

In the previous formulae, $u_{\rm max}=\mbox{arcsinh}
(z_{\rm max}/\xi)$.
 
\end{document}